\documentclass[journal]{IEEEtran}

\usepackage{stmaryrd}
\usepackage{amsmath}
\usepackage{amssymb}
\usepackage{amsthm}
\usepackage{slashbox,pict2e}
\usepackage{xcolor}
\usepackage{float}
\usepackage{stfloats}
\usepackage{graphicx}
\usepackage{stmaryrd}
\usepackage{lineno,hyperref}
\modulolinenumbers[5]
\usepackage{epstopdf}
\usepackage{graphicx,cite,multirow,rotating}
\usepackage{setspace}
\usepackage{pgfplots} 
\usepackage{tikz}
\usepackage{comment}
\usepackage{enumitem}   
\usepackage{lipsum}
\newtheorem*{Remark}{Remark}

\usetikzlibrary{shapes,arrows}
\usepackage{varwidth}
\usetikzlibrary{shapes.geometric, arrows}
\usetikzlibrary{positioning,arrows}

\usepackage{amsfonts}
\usepackage{yfonts}
\usepackage{psfrag}
\usepackage{pifont}
\usepackage{amsfonts}
\usepackage{bbm}
\usepackage{dsfont}

\usepackage{color}
\usepackage{amsmath,stackrel,dsfont}
\usepackage{amssymb,hyperref,stmaryrd}
\usepackage{tabularx}
\usepackage{subcaption}
\usepackage{booktabs}
\usepackage{algorithm}
\usepackage{algpseudocode}
\usepackage{pifont}
\modulolinenumbers[5]
\usepackage{amsmath}
\usepackage{epstopdf}
\usepackage{graphicx,cite,multirow,rotating}
\usepackage{setspace}
\usepackage{pgfplots} 
\usepackage{tikz}
\usepackage{flushend}
\usepackage{enumitem}

\usetikzlibrary{shapes,arrows}
\usepackage{varwidth}
\usetikzlibrary{shapes.geometric, arrows}
\usetikzlibrary{positioning,arrows}
\usepackage{amssymb}
\usepackage{amsfonts}
\usepackage{yfonts}
\usepackage{psfrag}
\usepackage{pifont}
\usepackage{amsfonts}
\usepackage{bbm}
\usepackage{dsfont}

\usepackage{color}
\usepackage{amsmath,stackrel,dsfont}
\usepackage{amssymb,hyperref,stmaryrd}
\usepackage{algorithm}
\usepackage{algpseudocode}
\usepackage{pifont}
\newcommand{\RN}[1]{%
	\textup{\uppercase\expandafter{\romannumeral#1}}%
}

\ifCLASSOPTIONcompsoc
\fi

\begin{document}
\title{DRL-Based Spectrum Sharing for RIS-Aided Local High-Quality Wireless Networks}

\author{Hamid~Reza~Hashempour,  Mina~Khadem, Eduard~A.~Jorswieck, \textit{Fellow, IEEE}, and Hien~Quoc~Ngo, \textit{Fellow, IEEE}
	\thanks{
   Hamid~Reza~Hashempour and Hien~Quoc~Ngo are with the Center for Wireless Innovation (CWI), Queen’s University Belfast, BT3 9DT Belfast, U.K., (Email:\{h.hashempoor, hien.ngo\}@qub.ac.uk).

    Mina Khadem is with the Department of Engineering,
    Universitat Pompeu Fabra (UPF), 08002 Barcelona, Spain (e-mail:
    mina.khadem@upf.edu).
    
    Eduard A. Jorswieck is with the Institute for Communication Technology, Technische Universitat Braunschweig, Germany (email: 
  e.jorswieck@tu-braunschweig.de).
	
    Eduard Jorswieck would like to thank the Federal Ministry of Research, Technology, and Space (BMFTR) for supporting the xG‑RIC project as part of the research program Communication Systems “Souverän. Digital. Vernetzt". (grant number 16KIS2429K).
    }}

\markboth{}%
{Shell \MakeLowercase{\textit{et al.}}: Bare Demo of IEEEtran.cls for IEEE Journals}

\setlength{\textfloatsep}{8.5pt plus 2pt minus 2pt}
\maketitle

\begin{abstract}
This paper investigates a smart spectrum-sharing framework for reconfigurable intelligent surface (RIS)-aided local high-quality wireless networks (LHQWNs) within a mobile network operator (MNO) ecosystem. Although RISs are often considered harmful due to interference, this work shows that properly controlled RISs can enhance quality of service (QoS). The proposed system enables temporary spectrum access for multiple vertical service providers (VSPs) by dynamically allocating radio resources. The spectrum is divided into dedicated subchannels assigned to individual VSPs and reusable subchannels shared among multiple VSPs, while RIS improves propagation conditions  and zero-forcing (ZF) precoding is adopted at the multi-antenna base station (BS) to cancel inter-user interference.
We formulate a multi-VSP utility maximization problem that jointly optimizes subchannel assignment, transmit power, and RIS phase configuration while accounting for spectrum access costs, RIS leasing costs, and QoS constraints. The resulting mixed-integer non-linear program (MINLP) is modeled as a Markov decision process (MDP) and solved using deep reinforcement learning (DRL). Deep deterministic policy gradient (DDPG) and soft actor--critic (SAC) algorithms are developed and compared. Numerical results show that SAC generally outperforms DDPG in convergence, stability, and utility, particularly in larger-scale scenarios. In the reduced-scale ablation study, both joint DRL methods outperform the  heuristic benchmark by at least $6.6\%$ in final moving-average reward.
\end{abstract}


\IEEEpeerreviewmaketitle

\section{Introduction}\label{intro}
\IEEEPARstart{W}{ith} the rapid growth of wireless communication networks,  spectrum scarcity has become a major challenge. Recent reports show that global mobile network data traffic grew by \(22\%\) between Q1 2025 and Q1 2026, driven by increasing demand for data-intensive services and emerging applications such as industrial connectivity, smart cities, and XR \cite{ericsson2026mobility}.
The emergence of vertical service providers (VSPs), which lease spectrum from mobile network operators (MNOs) to deploy local high-quality wireless networks (LHQWNs), has been proposed as a solution to improve spectral efficiency and service customization. 
However, traditional spectrum allocation schemes lack flexibility, leading to inefficient spectrum utilization. To address this issue, licensed shared access (LSA) and its evolution, evolved LSA (eLSA), have been proposed to enable controlled and dynamic spectrum sharing between MNOs and VSPs \cite{etsi2020elsa}.
In the eLSA framework, spectrum resources are categorized into dedicated subchannels, allocated exclusively to a single VSP, and reusable subchannels, which can be shared among multiple VSPs simultaneously. The MNO is responsible for dynamically assigning spectrum resources to VSPs based on their demand and network conditions. However, interference among VSPs using reusable subchannels poses a major challenge, impacting quality of service (QoS) \cite{patil2024spectrum}. 

To enhance network performance, reconfigurable intelligent surfaces (RISs) have emerged as a promising technology. RISs can manipulate the wireless propagation environment to improve coverage, mitigate interference, and enhance spectral efficiency \cite{hashempour2}. 
In the proposed framework, RISs are integrated into the eLSA ecosystem and can be leased by VSPs to satisfy application-specific QoS requirements through joint optimization. 

To efficiently allocate resources, we formulate a utility maximization problem for VSPs, taking into account the costs associated with leasing subchannels and RIS elements, power consumption, and the revenue generated from the profit per transmitted sum rate by dimension (\$/Mbps) for VSP $v$. 
Since utility revenue depends on users' achievable rates, QoS is captured through a minimum-rate constraint. However, the formulated problem is a non-convex mixed-integer nonlinear programming (MINLP) model, which is difficult to solve due to interdependencies between subchannel allocation, base station (BS) power control, and RIS configuration \cite{xu2024intelligent}.

To address the above challenges, we propose deep reinforcement learning (DRL)-based frameworks for dynamic spectrum sharing in RIS-aided local high-quality wireless networks. The considered resource allocation problem is first modeled as a Markov decision process (MDP), which captures the sequential and coupled nature of spectrum assignment, transmit power control, and RIS configuration. To tackle the resulting high-dimensional and hybrid continuous–discrete action space, we employ two representative DRL algorithms, namely deep deterministic policy gradient (DDPG) and soft actor--critic (SAC), with SAC generally achieving improved learning stability and higher utility, particularly in larger-scale scenarios.
The main contributions of this work are summarized as follows:
\begin{itemize}
    \item We propose a utility-driven RIS-assisted eLSA spectrum-sharing framework for a multi-VSP wireless ecosystem. The proposed model goes beyond conventional spectrum-sharing designs by incorporating economic utility as the main performance metric, jointly capturing the revenue from user service, the cost of spectrum and RIS leasing, transmit-power expenditure, and QoS satisfaction.

    \item We formulate an interference-aware multi-cell multiple-input single-output (MISO) utility maximization problem, where zero-forcing (ZF) precoding is adopted at the BSs to suppress intra-cell interference. This allows the resource-allocation design to focus on user scheduling, transmit-power allocation, and RIS phase configuration over dedicated and reusable subchannels, while still capturing inter-cell and inter-VSP interference.

    \item We develop a constraint-aware DRL solution framework by modeling the problem as an MDP and designing feasible action-mapping mechanisms for the mixed discrete--continuous resource-allocation variables.

   \item We tailor and compare DDPG and SAC for the proposed problem. Numerical results show that SAC generally achieves faster convergence, higher utility, and improved learning stability compared with DDPG. In the reduced-scale ablation study, both joint DRL methods outperform the exhaustive discrete
search (EDS) followed by alternating optimization (AO) heuristic benchmark, improving the final moving-average reward by at least $6.6\%$.
\end{itemize}

\subsection{Literature review}
The existing literature can be broadly categorized into three distinct sections, reflecting the comprehensive scope and diverse topics addressed in this paper: 1) Spectrum sharing for local high-quality wireless networks (LHQWNs), 2) RIS-assisted networks, 3) DRL for wireless resource management.

\subsubsection{Spectrum sharing for LHQWNs}
LSA and its enhanced form, eLSA, provide a regulated spectrum-sharing paradigm in which spectrum resources can be dynamically leased while preserving predictable service quality and interference protection. This paradigm is particularly relevant for LHQWNs, where private, non-public, and multi-tenant deployments require reliable spectrum access, resource isolation, and efficient coordination among multiple stakeholders. 
Recent studies have investigated different aspects of LSA/eLSA and local spectrum sharing. In \cite{khadem2024dynamic}, an eLSA framework is proposed that combines auction-based spectrum allocation, UAV-assisted sensing, and DRL to improve fairness and spectral efficiency among mobile network operators. In \cite{onidare2023optimizing}, an optimization framework is developed for LSA systems to jointly improve spectral efficiency and energy efficiency during incumbent spectrum usage. In \cite{khadem2024ai}, a QoS-aware spectrum management framework is proposed for beyond-5G and 6G systems, where verticals lease spectrum from an MNO through auction mechanisms and DRL is used to learn efficient allocation policies under dynamic conditions. More recently, utility-aware and incentive-driven local spectrum coordination mechanisms have also been studied in \cite{basaure2025utility,mu2025compete}, showing the importance of explicitly modeling spectrum-holder utility and coordination among neighboring local networks.
However, the above works mainly focus on auction design, spectrum allocation, energy/spectral efficiency, or interference coordination, while the joint role of RIS-assisted propagation control, reusable/dedicated subchannel assignment, power allocation, QoS constraints, and spectrum/RIS leasing costs in an eLSA-based multi-VSP ecosystem remains less explored.

\subsubsection{RIS-aided networks} RISs are emerging as an energy-efficient approach to enhance spectral efficiency and QoS in future wireless networks. By adaptively configuring the phase of reflected signals, RISs enable passive beamforming that strengthens desired signals and can suppress interference with low hardware cost and convenient integration into existing infrastructure \cite{ chen2021qos,Hashempour1,hashempour3}. 
Recent RIS-assisted designs have shown that joint optimization of active beamforming, power allocation, and RIS phase shifts can significantly improve the achievable rate of MISO/MIMO systems while satisfying QoS constraints such as minimum-rate or SINR requirements \cite{choi2024wmmse,alqwider2024ris}. Moreover, recent studies confirm that RISs can reshape wireless propagation to mitigate blockage, path loss, and fading by creating alternative or virtual LoS links, which is particularly useful in coverage-limited scenarios \cite{nwufo2025ris}. RIS-assisted spectrum sharing has also been recently studied in underlay cognitive radio networks, where the secondary network shares licensed spectrum while satisfying primary-network interference constraints \cite{lin2022intelligent,albadarneh2025performance}. In addition, RIS-assisted cooperative spectrum sensing has been proposed to enhance primary-signal reception and improve sensing reliability in cognitive radio networks \cite{xu2025drl}. However, to the best of our knowledge, the impact of RIS on utility maximization in a multi-VSP ecosystem has not been investigated.

\subsubsection{DRL for wireless resource management}
Wireless resource management problems, including spectrum sharing, dynamic spectrum access (DSA), power control, and scheduling, are typically time-varying, coupled across users, and difficult to solve optimally in real time. DRL has therefore been widely used to learn resource-control policies directly from interaction data. In spectrum sharing and DSA, DRL-based methods have been developed to adapt spectrum access decisions to uncertain traffic and interference conditions, as in heterogeneous-agent DSA for cognitive wireless networks \cite{wang2025heterogeneous}. Beyond access-only decisions, DRL has also been applied to shared-spectrum licensing and assignment, where a centralized agent jointly allocates spectrum and related resources based on real-time demand \cite{atimati2025resource}.
For LHQWNs, DRL has been combined with economic mechanisms to support spectrum leasing and service guarantees. For example, \cite{khadem2024ai} studies auction-based spectrum management where verticals lease spectrum from an MNO under minimum service constraints, while \cite{eldeeb2025offline} develops an offline multi-agent reinforcement learning framework for radio resource management that improves both sum-rate and tail-rate performance. 
DRL has also been applied to RIS-assisted resource management, where learning agents jointly adapt RIS configurations and communication resources in dynamic environments, including mobile multi-user MISO systems, cooperative spectrum sensing, and vehicular networks \cite{alqwider2024ris,xu2025drl,wang2025deep}.

Table~\ref{tab:comparison} compares the proposed framework with representative related works. Unlike these studies, our work jointly considers RIS-assisted transmission, multi-antenna ZF-based MISO downlink, reusable/dedicated subchannel assignment, transmit-power control, QoS constraints, and spectrum/RIS leasing costs for utility maximization in a multi-VSP eLSA ecosystem.
\begin{table}[t]
\caption{Comparison with Related Works}
\label{tab:comparison}
\centering
\scriptsize 
\setlength{\tabcolsep}{2.4pt}
\renewcommand{\arraystretch}{1.05}
\resizebox{\columnwidth}{!}{
\begin{tabular}{|l|c|c|c|c|c|c|c|}
\hline
\textbf{Feature}
& \textbf{Ours}
& \cite{khadem2024dynamic,khadem2024ai}
& \cite{basaure2025utility}
& \cite{xu2025drl}
& \cite{alqwider2024ris,wang2025deep}
& \cite{wang2025heterogeneous,atimati2025resource,eldeeb2025offline}
& \cite{Hashempour1,hashempour3} \\
\hline
LSA/eLSA                  & \checkmark & \checkmark &  &  &  &  &  \\ \hline
LHQWNs            & \checkmark & \checkmark & \checkmark &  &  &  &  \\ \hline
Multi-VSP                     & \checkmark & \checkmark & \checkmark &  &  &  &  \\ \hline
RIS-assisted    & \checkmark &  &  & \checkmark & \checkmark &  & \checkmark \\
\hline
MISO     & \checkmark &  &  &  & \checkmark &  & \checkmark \\ \hline
Resource allocation                & \checkmark & \checkmark &  &  & \checkmark & \checkmark & \checkmark \\ \hline
DRL-based              & \checkmark & \checkmark &  & \checkmark & \checkmark & \checkmark &  \\ \hline
Spectrum sharing & \checkmark & \checkmark & \checkmark & \checkmark &  & \checkmark &  \\ \hline
Utility modeling              & \checkmark & \checkmark & \checkmark &  &  &  &  \\ \hline
QoS-aware              & \checkmark &  &  &  & \checkmark &  &  \\ \hline
RIS leasing                   & \checkmark &  &  &  &  &  &  \\
\hline
\end{tabular}
}
\end{table}
\subsection{Organization and Notation}
The remainder of this paper is structured as follows: Section~\ref{Sys_Model} presents the system model and problem formulation. Section~\ref{sec:DRL} details the proposed DRL-based  framework. Section~\ref{Simulat} provides numerical results and performance evaluation. Finally, Section~\ref{conc} concludes the paper and outlines future research directions. 

\textit{Notation}: We use bold lowercase letters for vectors and bold uppercase letters for matrices. The notation $(\cdot)^T$ and $(\cdot)^H$ denote the transpose operator and the conjugate transpose operator, respectively. The symbol $\triangleq$ denotes a definition. The sets $\mathbb{R}^{N}$ and $\mathbb{C}^{N}$ represent real and complex $N$-dimensional vectors, respectively. $\mathcal{CN}(0,\sigma^2)$ denotes a complex circularly symmetric Gaussian random variable with variance $\sigma^2$.The operator $\mathrm{diag}\{\cdot\}$ constructs a diagonal matrix from its vector argument, and $[\mathbf{x}]_m$ denotes the $m$-th element of vector $\mathbf{x}$.

\section{System Model and Problem Formulation}
\label{Sys_Model}

\subsection{System Model}

\begin{figure} 
	\centering
	\includegraphics[width=1\linewidth]{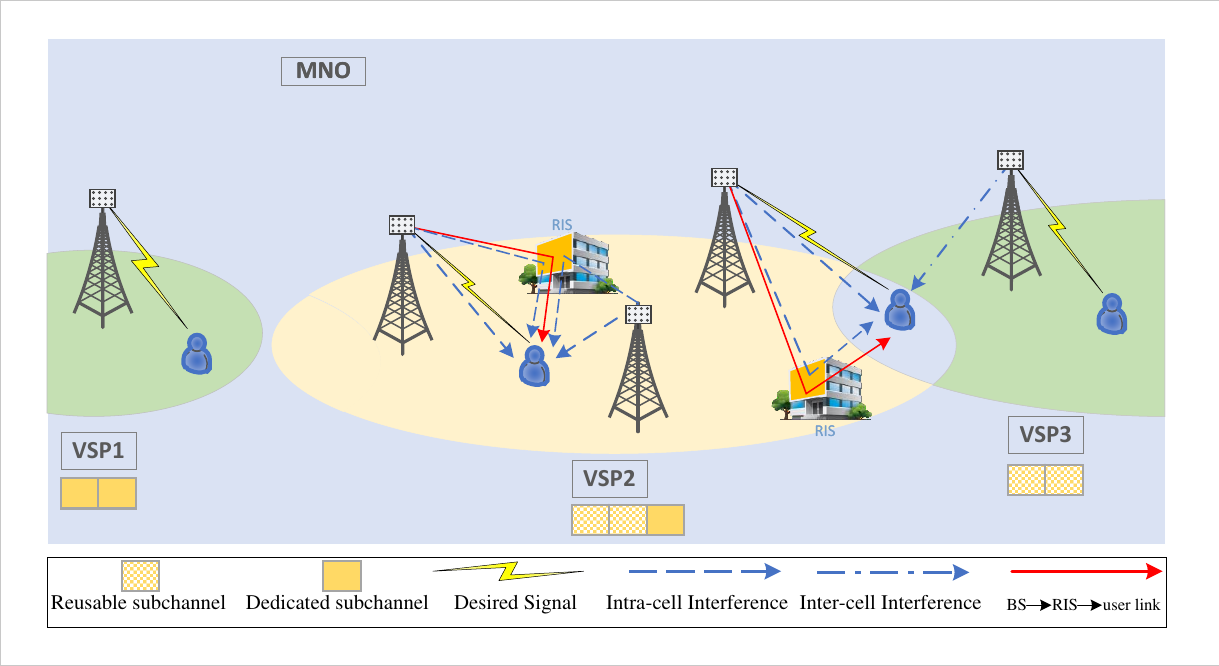}
	\caption{System model of an RIS-assisted multi-VSP wireless network within an MNO ecosystem with dedicated and reusable subchannels.}
	\label{fig2}		
\end{figure}

As shown in Fig.~\ref{fig2}, we consider an RIS-assisted LHQWN within an MNO ecosystem. The MNO acts as the spectrum owner and
resource coordinator, and dynamically allocates part of its licensed spectrum to
multiple VSPs operating in localized service areas.
This setting is consistent with the eLSA
framework, where spectrum resources can be assigned to local service providers
under predefined service areas, spectrum-usage rules, and QoS requirements
\cite{etsi2020elsa}.

In the considered network, the available spectrum is divided into dedicated and
reusable subchannels. Dedicated subchannels are exclusively assigned to specific
VSPs to guarantee reliable service, whereas reusable subchannels may be shared
among different VSPs, subject to interference control. To improve link quality and
mitigate inter-VSP interference, RISs are deployed to assist the transmissions.



Let $\mathcal{V} = \{1,2,\ldots,v,\ldots,V\}$ denote the set of VSPs in the MNO domain. For each VSP $v \in \mathcal{V}$, the set $\mathcal{B}_v = \{1,2,\ldots,b_v,\ldots,B_v\}$ denotes the BSs serving its users, while $\mathcal{K}_v = \{1,2,\ldots,k_v,\ldots,K_v\}$ represents the corresponding set of users. 
Moreover, $\mathcal{B} = \{\mathcal{B}_1, \mathcal{B}_2, \ldots, \mathcal{B}_V\}$ denotes the collection of BS sets associated with all VSPs. 
The set of all users are denoted by $\mathcal{K}= \{\mathcal{K}_{1}, \mathcal{K}_{2}, \cdots, \mathcal{K}_{v}, \cdots, \mathcal{K}_{V}\}$. 
We consider a downlink MISO transmission scenario, where each BS is equipped with $N_t$
 transmit antennas, while each user is equipped with a single antenna.
In order to improve the rate of users in the VSPs, a set of 
$\mathcal{J}= \{1, 2, \cdots, j, \cdots, J\}$ RISs is utilized. The RISs are parts of the MNO that are used by some VSPs to tackle QoS requirements of their users based on their applications.

We consider an enhanced LSA based spectrum sharing method, where the spectrum is allocated to VSPs based on their demand. We consider a set of $\mathcal{
C} = \{1,2, \cdots,c, \cdots, C\}$  available orthogonal subchannels for the MNO to share with VSPs. We define a set of $\mathcal{C}^d_v$ for the dedicated subchannels to be assured agreed level of QoS of each VSP and a set of $\mathcal{C}^r_v$ for reusable subchannels of each VSP  if the location areas of VSPs do not overlap or the MNO can handle interference where $\mathcal{C}^{r}_v,\mathcal{C}^d_v \subset \mathcal{C}$.
In addition, we define a binary indicator variable $\delta_c$, which equals $1$ if subchannel $c$ is reusable (i.e., $c \in \mathcal{C}_v^{r}$) and can be shared among VSPs, and $0$ otherwise.
The bandwidth of all subchannels are identical, and is denoted as $B_c$.
Let $\omega_{k,v}^{b,c}$ be the downlink binary subchannel assignment indicator of user $k$ served by BS $b$ of VSP $v$ over subchannel $c$, which is defined as follows
\begin{align}
\omega_{k,v}^{b,c}=
\begin{cases}
1, & \text{if subchannel $c$ is assigned to user $k$ in} \\ & \text{
	BS $b$ of VSP $v$};\\
0, & \text{Otherwise}.
\end{cases}
\end{align}
Subchannel $c$ can not be assigned to more than
$L_c$ users in the coverage of one BS, simultaneously. Therefore we introduce the following subchannel allocation constraint:
\begin{align} \label{eq:subchannel_constraint}
\sum_{k \in\mathcal{K}_{v}} \omega_{k,v}^{b,c} \leq L_c,
\quad \forall v\in\mathcal{V},~\forall b\in\mathcal{B}_v,~\forall c\in\mathcal{C}.
\end{align}
Let $p_{k,v}^{b,c}\ge 0$ denote the transmit power allocated by BS \(b\) of VSP \(v\) to user \(k\) over subchannel \(c\). The corresponding transmit beamforming vector is expressed as
\begin{align}
\mathbf{w}_{k,v}^{b,c}=\sqrt{p_{k,v}^{b,c}}\mathbf{f}_{k,v}^{b,c},
\end{align}
where $\mathbf{f}_{k,v}^{b,c}\in\mathbb{C}^{N_t\times 1}$ is the unit-norm ZF beamforming direction, i.e., $\left\Vert \mathbf{f}_{k,v}^{b,c}\right\Vert=1$. The per-BS transmit power constraint is
\begin{align}
\sum_{c\in\mathcal{C}}\sum_{k\in\mathcal{K}_v} p_{k,v}^{b,c}\le P_{\max}^{b,v},
\quad \forall v\in\mathcal{V},~\forall b\in\mathcal{B}_v,
\label{eq:power_budget}
\end{align}
and power is active only when the user is scheduled
\begin{align}
0\le p_{k,v}^{b,c}\le \omega_{k,v}^{b,c} P_{\max}^{b,v},
\quad \forall v,b,k,c.
\label{eq:power_link}
\end{align}
Considering that different BSs may serve different sets of users, we define the binary BS-association indicator $\varphi_{k,v}^{b}\in\{0,1\}$, where $\varphi_{k,v}^{b}=1$ if user $k\in\mathcal{K}_v$ is associated with BS $b\in\mathcal{B}_v$ of VSP $v$, and $\varphi_{k,v}^{b}=0$ otherwise. Each user can be associated with at most one BS at any time, i.e.,
\begin{align}
\sum_{b\in \mathcal{B}_{v}} \varphi_{k,v}^{b} \le 1, 
\quad \forall v\in\mathcal{V},~\forall k\in \mathcal{K}_v .
\label{eq:bs_assoc}
\end{align}
Moreover, each user can be scheduled on at most one subchannel from its associated BS. This constraint is enforced by
\begin{align}
\sum_{b\in\mathcal{B}_v}\sum_{c\in\mathcal{C}} \omega_{k,v}^{b,c} \le 1,
\quad \forall v\in\mathcal{V},~\forall k\in\mathcal{K}_v .
\label{eq:one_channel}
\end{align}
Finally, scheduling is only allowed if the corresponding BS association holds, i.e.,
\begin{align}
\omega_{k,v}^{b,c} \le \varphi_{k,v}^{b},
\quad \forall v\in\mathcal{V},~\forall b\in\mathcal{B}_v,~\forall k\in\mathcal{K}_v,~\forall c\in\mathcal{C}.
\label{eq:link_assoc_sched}
\end{align}
The reflection-coefficient matrix of the $j$th RIS is defined as
\begin{align}
\mathbf{\Theta}_j \triangleq \mathrm{diag}\!\left(e^{\mathrm{j}\theta_{j,1}}, e^{\mathrm{j}\theta_{j,2}}, \ldots, e^{\mathrm{j}\theta_{j,M_j}}\right),
\quad ~\forall m\in\mathcal{M}_j,
\label{RIS}
\end{align}
where $\theta_{j,m}\in[0,2\pi)$. Furthermore, we define $d_{j}^{k}$ as a binary indicator denoting whether
user $k$ lies within the effective coverage region of RIS $j$, where
$d_{j}^{k}=1$ if RIS $j$ can assist user $k$, and $0$ otherwise.
Since a user is typically located within the dominant coverage region of
its nearest RIS, each user is assumed to be associated with at most one
RIS, i.e.,
\begin{align}
\sum_{j\in\mathcal J} d_j^k \le 1,\quad \forall k\in\mathcal K.
\end{align}
 The channel coefficients from BS $b$ to user $k$, from RIS $j$ to user $k$, and from BS $b$ to RIS $j$ on subchannel $c$ are denoted by $\mathbf{h}_{b,k}^{c}\in\mathbb{C}^{N_t\times 1}$, $\mathbf{r}^c_{j,k} \in \mathbb{C}^{M_j \times 1}$, and $\mathbf{G}_{b,j}^{c}\in\mathbb{C}^{M_j\times N_t}$, respectively. Let $\mathcal{U}_{b,c}$ denote the set of users simultaneously scheduled by BS $b$ on subchannel $c$, where $|\mathcal{U}_{b,c}|=U_{b,c}\le L_c\le N_t$. The effective RIS-assisted channel matrix is constructed as
\begin{align}
\tilde{\mathbf H}_{b,c}
=
\left[
\tilde{\mathbf h}_{b,1}^{c},
\tilde{\mathbf h}_{b,2}^{c},
\ldots,
\tilde{\mathbf h}_{b,U_{b,c}}^{c}
\right]
\in\mathbb{C}^{N_t\times U_{b,c}},
\end{align}
where $
\tilde{\mathbf h}_{b,k}^{c}
\triangleq
\mathbf h_{b,k}^{c}
+
\sum_{j\in\mathcal J}
d_j^k
(\mathbf G_{b,j}^{c})^H
\mathbf\Theta_j
\mathbf r_{j,k}^{c}$
is the effective RIS-assisted channel between BS $b$ and user $k$. The corresponding ZF beamforming matrix is obtained as
\begin{align}
\mathbf F_{b,c}
=
\tilde{\mathbf H}_{b,c}
\left(
\tilde{\mathbf H}_{b,c}^{H}
\tilde{\mathbf H}_{b,c}
\right)^{-1},
\end{align}
where the unit-norm beamforming direction of user $k$ is obtained by normalizing the corresponding column of $\mathbf F_{b,c}$, i.e., $\mathbf f_{k,v}^{b,c}
=
\mathbf F_{b,c}(:,k) /
\left\|
\mathbf F_{b,c}(:,k)
\right\|$.
Accordingly, under perfect channel state information (CSI) and provided that the effective channel
matrix $\tilde{\mathbf H}_{b,c}$ has full column rank, the normalized ZF beamformers satisfy
\begin{align}
(\tilde{\mathbf h}_{b,k}^{c})^{H}
\mathbf f_{u,v}^{b,c}
=
0,\quad \forall u\neq k,
\end{align}
thereby suppressing intra-cell interference among the users
co-scheduled by the same BS on the same subchannel.
Then, the received interference at user $k$, associated with BS $b$ of VSP $v$ on subchannel $c$, is expressed as
$ I_{k,v}^{b,c} =  I_1 + I_2 $, where
\begin{align}
I_1 &\triangleq 
\sum_{\substack{b'\in\mathcal{B}_{v}\\ b'\neq b}}
\sum_{u\in\mathcal{K}_v}
\omega_{u,v}^{b',c}p_{u,v}^{b',c}\,\big|(\tilde{\mathbf{h}}_{b',k}^{c})^H \mathbf{f}_{u,v}^{b',c}\big|^2
\end{align}
represents the intra-VSP interference, and
\begin{align}
I_2 &\triangleq 
\delta_c
\sum_{\substack{v'\in\mathcal{V}\\ v'\neq v}}
\sum_{b'\in\mathcal{B}_{v'}}
\sum_{u\in\mathcal{K}_{v'}}
\omega_{u,v'}^{b',c}p_{u,v'}^{b',c}\,\big|(\tilde{\mathbf{h}}_{b',k}^{c})^H \mathbf{f}_{u,v'}^{b',c}\big|^2
\end{align}
corresponds to the inter-VSP interference.
It is worth noting that an RIS is a passive reflecting element and does not actively generate interference. In this work, we therefore consider the RIS-reflected components of both the desired and interfering signals propagating  through the BS–RIS–user cascaded links.
\begin{Remark}
Each RIS is assumed to be deployed and controlled by its geographically nearest BS; hence, the RIS--BS association is fixed by the network topology and is not optimized. This is consistent with practical deployments, where each RIS is connected to a single BS controller via a wired or wireless control link. The BS--RIS control links are assumed to be ideal, i.e., RIS phase-control commands are delivered without delay, errors, or signaling overhead. The impact of imperfect control signaling is left for future work. Due to severe path loss, blockage, and cascaded double fading, links between users and non-associated RISs are considered negligible. Therefore, each user can benefit from at most one RIS, and cross-RIS reflections are not included in the received signal model.
\end{Remark}
The received signal-to-interference-plus-noise ratio (SINR) at user $k$ from the $b$th BS over subchannel $c$ to decode its own signal which is denoted by $\gamma_{k,v}^{b,c}$ is obtained as
\begin{align}	
\gamma_{k,v}^{b,c}
=
\frac{\omega_{k,v}^{b,c} \, p_{k,v}^{b,c}
\big|(\tilde{\mathbf{h}}_{b,k}^{c})^H \mathbf{f}_{k,v}^{b,c}\big|^2}
{I_{k,v}^{b,c}+B_c N_0},
 \ \forall v, b, k, c,
\label{eq:SINR_clean}
\end{align}
where $N_0$ stands for the power spectral
density of noise.
The corresponding achievable data rate is
\begin{align}
R_{k,v}^{b,c} = B_c \log_2\!\left(1+ \gamma_{k,v}^{b,c}\right).
\end{align}
Thus, the total rate of the $k$th user is
\begin{align}
R_{k,v} = \sum_{b \in \mathcal{B}_v} \sum_{c \in \mathcal{C}} R_{k,v}^{b,c},
\quad \forall k\in \mathcal{K}_v,~\forall v\in \mathcal{V}.
\end{align}
Consider that all users want to obtain their maximum transmission capacity while meeting a
minimum QoS requirement $R_{k,v}^{th}$. Thus, we enforce that the
rate of the $k$th user $R_{k,v}$ should be not less than the minimum
QoS requirement $R_{k,v}^{th}$.

\subsection{Problem Formulation}
We aim to maximize the utility of the VSPs, where the utility of each VSP consists of a revenue function and a cost function. In the following parts, we formulate the revenue, cost, and utility functions, respectively.
\\\indent $\bullet $ \textbf{Cost Function:}
As part of our system model, we take into account four types of costs: reusable and dedicated subchannels, RIS, and transmitted power. Accordingly, the total cost function of each VSP is denoted by $\mathbb{U}_{v}^{\text{Cost}}$ and defined as 
\begin{align}\label{cost-func}
\mathbb{U}_{v}^{\text{Cost}}
&=
\underbrace{N_{v}^r \lambda^r + N_{v}^d \lambda^d}_{\mathrm{Cost\ of\ spectrum}}
+
\underbrace{N_{v}^j \psi^j}_{\mathrm{Cost\ of\ RIS}} \nonumber\\
&\quad+
\underbrace{\alpha_p B_c
\sum_{b \in \mathcal{B}_v}
\sum_{k \in \mathcal{K}_v}
\sum_{c \in \mathcal{C}}
p_{k,v}^{b,c}}_{\mathrm{Cost\ of\ transmitted\ power}},
\quad \forall v \in \mathcal{V}.
\end{align}
where the $N_{v}^r$, $N_{v}^d$ and $N_{v}^j$ are the number of reusable subchannels, dedicated subchannels and used RISs for transmission, respectively. These quantities are known to both the VSPs and the MNO. Let $\lambda^r > 0$, $\lambda^d > 0$ and $\psi^j > 0$ represent the price of each reusable subchannel, price of each dedicated subchannel and price of each RIS leasing, respectively.   Considering that RISs belong to the MNO and $\alpha_p>0$ represents the unit price of the transmitted power (with unit \$/Watt/Hz).
\\\indent $\bullet $ 
\noindent\textbf{Revenue Function:}
Let $\beta_v>0$ denote the profit of VSP $v$ per unit transmitted data
rate (with unit \$/Mbps), corresponding to a linear usage-based pricing
model.
 We denote the revenue function of each VSP by $\mathbb{U}_{v}^{\text{Revenue}}$. Accordingly, it can be formulated as follows
\begin{align}
\mathbb{U}_{v}^{\text{Revenue}}
=
\beta_v
\sum_{b \in \mathcal{B}_v}
\sum_{k \in \mathcal{K}_v}
\sum_{c \in \mathcal{C}}
R_{k,v}^{b,c}
=
\beta_v R_v,
\quad \forall v \in \mathcal{V}.
\end{align}
\\\indent $\bullet $ 
\noindent\textbf{Utility Function:}
The utility function of VSP $v$ is defined as the difference between its revenue and cost. As a result, it can be calculated as follows
\begin{align}
	\mathbb{U}_v = \Phi_1{\mathbb{U}_{v}^{\text{Revenue}}} - \Phi_2{\mathbb{U}_{v}^{\text{Cost}}}, \forall v \in \mathcal{V},
\end{align}
where $\Phi_1,\Phi_2>0$ are scaling factors used to balance the contributions of the revenue and cost terms in the utility function.
Our objective is to jointly optimize the subchannel allocation, BS association, RIS phase control, and transmit power allocation so as to maximize the overall utility of the VSPs, while guaranteeing the QoS requirements of all users. Mathematically, the utility maximization problem for all VSPs is formulated as follows
\begin{subequations} \label{P1}
\allowdisplaybreaks
\begin{align}
\max_{\boldsymbol{\omega}, \mathbf{p}, \boldsymbol{\varphi}, \boldsymbol{\theta}}
&\sum_{v\in\mathcal{V}} \mathbb{U}_v \label{P1-a}\\
\mathrm{s.t.}\quad
&R_{k,v} \ge R_{k,v}^{\mathrm{th}},
\quad \forall v\in\mathcal{V},~\forall k\in\mathcal{K}_v, \label{P1-b}\\
&\omega_{k,v}^{b,c},\varphi_{k,v}^{b} \in\{0,1\},
\quad \forall v,b,k,c, \label{P1-j}\\
&\theta_{j,m}\in[0,2\pi),
\quad \forall j\in\mathcal{J},~\forall m\in\mathcal{M}_j. \label{P1-k}
\\&\text{\eqref{eq:subchannel_constraint}--\eqref{eq:link_assoc_sched}.}
\end{align}
\end{subequations}
The boldface symbols denote the collections of the corresponding optimization variables, defined as

\begin{align}
\allowdisplaybreaks
\boldsymbol{\omega} &\triangleq \{\omega_{k,v}^{b,c} \mid \forall k\in\mathcal{K}_v,~\forall v\in\mathcal{V},~\forall b\in\mathcal{B}_v,~\forall c\in\mathcal{C}\}, \\
\mathbf{p} &\triangleq \{p_{k,v}^{b,c} \mid \forall k\in\mathcal{K}_v,~\forall v\in\mathcal{V},~\forall b\in\mathcal{B}_v,~\forall c\in\mathcal{C}\}, \\
\boldsymbol{\varphi} &\triangleq \{\varphi_{k,v}^{b} \mid \forall k\in\mathcal{K}_v,~\forall v\in\mathcal{V},~\forall b\in\mathcal{B}_v\}, \\
\boldsymbol{\theta} &\triangleq \{\theta_{j,m} \mid \forall j\in\mathcal{J},~\forall m\in\mathcal{M}_j\}.
\end{align}
Moreover, constraint \eqref{P1-j} ensures that the corresponding decision variables are binary. The proposed problem formulation \eqref{P1} is an MINLP problem, which is difficult to solve in polynomial time. Moreover, the subchannel allocation, BS association, and power control strategies of each VSP are strongly coupled due to mutual interference. In addition, the dynamic wireless channel conditions and time-varying network environment further complicate the problem, making it challenging to solve using conventional optimization methods. These challenges motivate the adoption of a DRL-based solution, as described in the next section.

\section{DRL-Based Solution} \label{sec:DRL}

In this section, we propose two DRL-based frameworks to solve the utility maximization problem~\eqref{P1}. 
Unlike conventional continuous-action
DRL formulations that optimize only continuous variables, the proposed
framework jointly addresses discrete scheduling decisions and continuous
resource allocation through a feasible action-mapping mechanism.
Specifically, we first model the joint optimization problem as a MDP. Then, we develop DRL solutions based on the DDPG and SAC algorithms, which are well suited for high-dimensional continuous control problems with coupled decision variables. These methods enable efficient learning of joint scheduling, power allocation, and RIS configuration policies under dynamic network conditions.

\subsection{MDP Formulation}\label{subsec:mdp}

We formulate the joint resource allocation problem as an MDP defined by the tuple
\begin{align}
\mathcal{M} \triangleq \langle \mathcal{S}, \mathcal{A}, \mathcal{P}, \mathcal{R} \rangle,
\end{align}
where $\mathcal{S}$ denotes the state space, $\mathcal{A}$ denotes the action space, $\mathcal{P}$ represents the state transition dynamics, and $\mathcal{R}$ is the reward function.

\subsubsection{State Space}

The state at time slot $t$, denoted by $s_t \in \mathcal{S}$, summarizes the essential information of the network environment required for sequential decision-making. It is defined as
\begin{align}\label{state_def}
s_t = \big[\mathbf{H}(t),~ \mathbf{R}(t),~ a_{t-1}\big],
\end{align}
where $\mathbf{H}(t)$ collects the instantaneous CSI of all communication links in the network, given by
\begin{align}
\mathbf{H}(t) \triangleq 
\Big\{ 
\mathbf{h}_{b,k}^c(t),~
\mathbf{G}_{b,j}^c(t),~
\mathbf{r}_{j,k}^c(t)
\;\big|\; \forall b,k,j,c
\Big\}.
\end{align}
Moreover, $\mathbf{R}(t)$ denotes the vector of achieved user data rates at time slot $t$, i.e.,
\begin{align}
\mathbf{R}(t) \triangleq \big\{ R_{k,v}(t) \mid \forall v\in\mathcal{V},~\forall k\in\mathcal{K}_v \big\}.
\end{align}
Finally, $a_{t-1}$ represents the previously executed feasible control action, including scheduling $\boldsymbol{\omega}$, transmit power $\mathbf{p}$, BS association $\boldsymbol{\varphi}$,  and RIS phase shifts $\boldsymbol{\theta}$. 
By incorporating the previous action, the state definition preserves the Markov property and enables the agent to capture the impact of past decisions on the current network dynamics.
\subsubsection{Action Space}

At each time step $t$, the agent selects a control action $a_t \in \mathcal{A}$, which jointly determines the scheduling, power allocation, and RIS configuration. The feasible action is defined as
\begin{align}\label{action_def}
a_t = \big[ \boldsymbol{\omega}(t),~ \mathbf{p}(t),~ \boldsymbol{\varphi}(t),~  \boldsymbol{\theta}(t) \big],
\end{align}
where $\boldsymbol{\omega}(t)$, and $\boldsymbol{\varphi}(t)$ denote discrete scheduling, and BS association variables, respectively, while $\mathbf{p}(t)$ and $\boldsymbol{\theta}(t)$ represent the continuous transmit power allocation and RIS phase shifts.

Since standard DRL algorithms operate over continuous action spaces, the actor network outputs a raw continuous action $\tilde a_t$, consisting of relaxed representations of the discrete variables and unconstrained continuous values. This raw action is subsequently mapped onto the feasible set $\mathcal{A}$ through deterministic projection, thresholding, and normalization operations. In particular, the relaxed binary variables are converted into feasible binary decisions using element-wise thresholding, i.e.,
\begin{equation}
x = 
\begin{cases}
1, & \text{if } \tilde{x} \geq 0.5,\\
0, & \text{otherwise},
\end{cases}
\end{equation}
while the continuous variables are clipped and rescaled to satisfy the corresponding box constraints.

\subsubsection{State Transition}

The state transition probability $\mathcal{P}(s(t+1)\mid s(t),a(t))$ is governed by the wireless channel evolution, user mobility, traffic dynamics, and the applied control actions. Since these dynamics are generally unknown and time-varying, a model-free DRL approach is adopted.

\subsubsection{Reward Function}

The immediate reward at time $t$ is designed based on the system utility and QoS satisfaction. It is defined as
\begin{align}\label{reward_def}
r(t) = \sum_{v\in\mathcal{V}} \mathbb{U}_v(t) - \lambda_{\mathrm{qos}}\sum_{v\in\mathcal{V}} \sum_{k\in\mathcal{K}_v} \max\big(0, R_{k,v}^{\mathrm{th}} - R_{k,v}(t)\big),
\end{align}
where the first term corresponds to the total utility of all VSPs, and the second term penalizes violations of QoS constraints with a weight $\lambda_{\mathrm{qos}}>0$.

\subsection{DDPG-Based Learning Framework}\label{subsec:ddpg}

To solve the MDP formulated in Section~\ref{subsec:mdp}, we adopt the DDPG algorithm, which is particularly suitable for high-dimensional continuous control problems with coupled decision variables. In our setting, the action space consists of continuous transmit powers and RIS phase shifts, as well as relaxed representations of discrete scheduling and association decisions, making DDPG a natural choice.

DDPG follows an actor--critic architecture, where the actor network learns a deterministic policy that maps the observed system state to a control action, while the critic network evaluates the quality of the selected action through a learned Q-function. By combining policy gradient updates with value-function approximation, DDPG enables stable learning in complex and nonconvex environments.

The objective of the learning process is to maximize the expected long-term discounted return
\begin{align}
\max_{\pi}~ \mathbb{E}\!\left[ \sum_{t=0}^{\infty} \gamma^t r(t) \right],
\label{eq:ddpg_obj}
\end{align}
where $r(t)$ is the instantaneous reward defined in \eqref{reward_def} and $\gamma \in (0,1)$ is the discount factor.

\subsubsection{Learning Procedure}

At each time step $t$, the actor network outputs a raw continuous action
\begin{align}
\tilde a_t = \pi(s_t;\psi),
\label{eq:ddpg_action}
\end{align}
where $\pi(\cdot)$ denotes the deterministic policy parameterized by $\psi$. 
As described in the MDP formulation, $\tilde a_t$ contains continuous relaxations of the hybrid decision variables. It is therefore mapped onto the feasible action set $\mathcal{F}$ via a deterministic projection operator
\begin{align}
a_t = \Pi_{\mathcal{F}}(\tilde a_t),
\label{eq:ddpg_proj}
\end{align}
which enforces all system constraints, including power budgets, scheduling feasibility, and RIS phase bounds. Specifically, the scheduling component is projected by assigning each user to a feasible subchannel at its deterministically associated BS while satisfying the subchannel multiplexing constraint. The power component is clipped and normalized to satisfy the per-BS transmit-power budget in \eqref{eq:power_budget}, while the RIS phase component is mapped to the feasible interval $[0,2\pi)$. This projection mechanism is  applied identically to both DDPG and SAC to ensure a fair comparison.

After executing $a_t$, the agent observes the reward $r_t = r(s_t,a_t)$ and the next state $s_{t+1}$. The transition tuple $(s_t,a_t,r_t,s_{t+1})$ is stored in the replay buffer $\mathcal{D}$.

The critic network $Q(s,a;\xi)$ is trained by minimizing the temporal-difference (TD) loss
\begin{align} \label{loss2}
L^{\mathrm{C}}_{\xi}
=
\mathbb{E}\!\left[
\Big(
r_t
+ \gamma Q^{\prime}(s_{t+1}, \pi^{\prime}(s_{t+1}); \xi^{\prime})
- Q(s_t,a_t;\xi)
\Big)^2
\right],
\end{align}
where $(\pi^{\prime}, Q^{\prime})$ denote the corresponding target actor and critic networks.
The actor network is updated by maximizing the critic’s output, which is equivalently formulated as minimizing the following surrogate loss:
\begin{align} \label{loss1}
L^{\mathrm{A}}_{\psi} = - Q\!\left(s_t, \pi(s_t;\psi); \xi \right).
\end{align}
Accordingly, the actor and critic parameters are updated via gradient descent as
\begin{align}
\psi &\leftarrow \psi - \eta_{\mathrm{A}} \nabla_{\psi} L^{\mathrm{A}}_{\psi},\\
\xi &\leftarrow \xi - \eta_{\mathrm{C}} \nabla_{\xi} L^{\mathrm{C}}_{\xi},
\end{align}
where $\eta_{\mathrm{A}}$ and $\eta_{\mathrm{C}}$ denote the learning rates of the actor and critic networks, respectively.
The corresponding target networks are softly updated using Polyak averaging
\begin{align}
\psi^{\prime} &\leftarrow \tau \psi + (1-\tau)\psi^{\prime}, \label{update1}\\
\xi^{\prime} &\leftarrow \tau \xi + (1-\tau)\xi^{\prime}, \label{update2}
\end{align}
where $\tau \in (0,1)$ is the soft update factor.

\subsection{SAC-Based Learning Framework}\label{subsec:sac}

To further enhance exploration efficiency and learning stability, we also adopt the SAC algorithm to solve the utility maximization problem~\eqref{P1}. SAC is an off-policy actor--critic method that incorporates an entropy-regularized objective, enabling robust learning in high-dimensional and nonconvex control problems. This property is particularly desirable in our setting, where the action space consists of continuous power variables, RIS phase shifts, and relaxed representations of discrete scheduling and association decisions.

Unlike DDPG, which learns a deterministic policy, SAC learns a stochastic policy that maximizes both the expected cumulative reward and the entropy of the policy. Specifically, the SAC objective is given by \cite{haarnoja2018soft}
\begin{align}
\max_{\pi}~ \mathbb{E}\!\left[\sum_{t=0}^{\infty} \gamma^t \Big(r(t) + \alpha \mathcal{H}(\pi(\cdot|s_t))\Big)\right],
\label{eq:sac_obj}
\end{align}
where $\mathcal{H}(\pi(\cdot|s_t))$ denotes the differential entropy of the policy at state $s_t$, and $\alpha>0$ is the temperature parameter controlling the tradeoff between reward maximization and exploration, which is automatically tuned during training.

\subsubsection{ Feasibility Projection (Environment Mapping)}
The feasibility projection is identical to that of the DDPG framework described in \eqref{eq:ddpg_proj}.
\subsubsection{Critic Update}
In the SAC framework, the actor network parameterized by $\psi$ defines a stochastic policy $\pi_{\psi}(\tilde a_t|s_t)$ over the raw continuous action. The raw action is sampled according to
\begin{align}
\tilde a_t \sim \pi_{\psi}(\cdot|s_t),
\label{eq:sac_sample}
\end{align}
and is then mapped to the feasible action $ a_t=\Pi_{\mathcal{F}}(\tilde a_t)$ via \eqref{eq:ddpg_proj}. After executing $a_t$, the environment returns the next state and reward according to the MDP in Section~\ref{subsec:mdp}. In particular, the instantaneous reward $r_t$ is computed using \eqref{reward_def}, where the achieved rates are obtained from the SINR expression in \eqref{eq:SINR_clean}.

To mitigate overestimation bias, SAC employs two critic networks $Q_1(s, a;\xi_1)$ and $Q_2(s, a;\xi_2)$ with target networks $Q_1'(\cdot;\xi_1')$ and $Q_2'(\cdot;\xi_2')$. For each transition $(s_t,a_t,r_t,s_{t+1})$, the soft target is defined as
\begin{align}
y_t &= r_t + \gamma\Big(\min_{i=1,2} Q_i'(s_{t+1}, a_{t+1};\xi_i') \nonumber\\
&\hspace{1.5em}- \alpha \log \pi_{\psi}(\tilde a_{t+1}|s_{t+1})\Big),
\label{eq:sac_target}
\end{align}
where $\tilde a_{t+1}\sim \pi_{\psi}(\cdot|s_{t+1})$ and $ a_{t+1}=\Pi_{\mathcal{F}}(\tilde a_{t+1})$. The critics are trained by minimizing the soft Bellman residual
\begin{align}
L^{\mathrm{C}}_{\xi_i}
=
\mathbb{E}\!\left[
\Big(Q_i(s_t,a_t;\xi_i)-y_t\Big)^2
\right],\quad i\in\{1,2\}.
\label{eq:sac_critic_loss}
\end{align}

\subsubsection{Actor and Temperature Updates}

The actor is updated using the reparameterization trick, where the raw action can be written as
\begin{align}
\tilde a_t
=
\tanh\!\left(
\boldsymbol{\mu}_{\psi}(s_t)
+
\boldsymbol{\sigma}_{\psi}(s_t)\odot\boldsymbol{\epsilon}
\right),
\quad
\boldsymbol{\epsilon}\sim\mathcal{N}(\mathbf{0},\mathbf I).
\end{align}
The corresponding feasible action is then obtained by \eqref{eq:ddpg_proj}. The actor network is updated by minimizing the entropy-regularized policy loss
\begin{align} 
L^{\mathrm{A}}_{\psi}
=
\mathbb{E}\!\left[
\alpha \log \pi_{\psi}(\tilde a_t|s_t)
-
\min_{i=1,2} Q_i(s_t,a_t;\xi_i)
\right].
\label{loss_sac_actor}
\end{align}
The first term encourages exploration, while the second term guides the policy toward actions with higher expected return.
Moreover, the temperature parameter $\alpha$ is adaptively adjusted during training by minimizing~\cite{haarnoja2018soft}
\begin{align}
L(\alpha)
=
\mathbb{E}\!\left[-\alpha \big(\log \pi_{\psi}(\tilde a_t|s_t) + \mathcal{H}_{\mathrm{target}}\big)\right],
\label{eq:sac_temp_loss}
\end{align}
where $\mathcal{H}_{\mathrm{target}}$ is a predefined target entropy. Finally, the target critic networks are softly updated as
\begin{align}
\xi_i' \leftarrow \tau \xi_i + (1-\tau)\xi_i', \quad i\in\{1,2\}.
\label{eq:sac_polyak}
\end{align}

By explicitly encouraging exploration through entropy regularization, SAC improves robustness and learning stability compared with deterministic policy-gradient methods. This makes it suitable for the considered high-dimensional hybrid resource-allocation problem involving scheduling, power allocation, and RIS phase configuration.

The complete DDPG/SAC-based learning
framework for solving Problem~\eqref{P1} is summarized in Algorithm~\ref{algorithm:drl}.

\vspace{-0.8em}
\begin{algorithm}[t]
\footnotesize
\caption{DDPG/SAC-Based Solution of Problem~\eqref{P1}}
\label{algorithm:drl}
\begin{algorithmic}[1]
\State Initialize replay buffer $\mathcal{D}$
\State Initialize actor, critic(s), and target network(s)
\If{SAC}
    \State Initialize $Q_1,Q_2$ and temperature $\alpha$
\EndIf

\For{episode $e=1,\ldots,E$}
\State Observe initial state $s_0$
\For{$t=0,\ldots,T-1$}

\If{DDPG}
    \State $\tilde a_t=\pi(s_t;\psi)$ by \eqref{eq:ddpg_action}
\Else
    \State Sample $\tilde a_t\sim\pi(\cdot|s_t;\psi)$ by \eqref{eq:sac_sample}
\EndIf

\State $a_t=\Pi_{\mathcal F}(\tilde a_t)$ by \eqref{eq:ddpg_proj}
\State Execute $a_t$, observe $(r_t,s_{t+1})$
\State Store $(s_t,a_t,r_t,s_{t+1})$ in $\mathcal D$
\State Sample a mini-batch from $\mathcal D$

\If{DDPG}
    \State Update critic by \eqref{loss2}
    \State Update actor by \eqref{loss1}
    \State Update target networks by \eqref{update1}--\eqref{update2}
\Else
    \State Compute targets by \eqref{eq:sac_target}
    \State Update critics by \eqref{eq:sac_critic_loss}
    \State Update actor by \eqref{loss_sac_actor}
    \State Update $\alpha$ by \eqref{eq:sac_temp_loss}
    \State Update target critics by \eqref{eq:sac_polyak}
\EndIf

\State $s_t\leftarrow s_{t+1}$

\EndFor
\EndFor
\end{algorithmic}
\end{algorithm}

\subsection{EDS With AO-Based Heuristic Refinement}
\label{subsec:eds_projected_ao}

We consider a two-stage benchmark comprising an EDS followed by an alternating numerical refinement of the
continuous power-allocation and RIS phase-shift variables. The BS association is fixed according to the minimum Euclidean distance criterion, whereas EDS enumerates all feasible subchannel-allocation
configurations $\boldsymbol{\omega}$ satisfying the scheduling
constraints, including the limit of at most $N_t$ scheduled users per
BS and subchannel.

For each feasible configuration, the transmit-power budget of each BS
is uniformly distributed among its active links. To ensure a fair
comparison, a common randomly initialized RIS phase vector is used for
all EDS candidates. The corresponding effective channels and ZF
beamforming vectors are then computed, and each candidate is evaluated
using the original SINR, rate, cost, utility, and QoS-penalty
expressions. The best configuration is selected according to the
QoS-aware comparison rule defined below. Let
$\boldsymbol{\omega}^{\ast}$ denote the selected scheduling
configuration, and let $\mathbf p^{(0)}$ and
$\boldsymbol{\theta}^{(0)}$ denote its initial power and RIS phase
vectors, respectively.

For fixed $\boldsymbol{\omega}^{\ast}$, the second stage alternately
refines $\mathbf p$ and $\boldsymbol{\theta}$ by numerically estimating
the local variation of the exact objective. Define the maximum QoS
shortfall as
\begin{align}
\Delta(\mathbf p,\boldsymbol{\theta})
=
\max_{\substack{v\in\mathcal V\\k\in\mathcal K_v}}
\left[
R_{k,v}^{\mathrm{th}}
-
R_{k,v}(\mathbf p,\boldsymbol{\theta})
\right]^+,
\label{eq:ao_max_qos_shortfall}
\end{align}
where $[x]^+=\max\{x,0\}$. The penalized objective is
\begin{align}
\mathcal F(\mathbf p,\boldsymbol{\theta})
=
\mathbb U(\mathbf p,\boldsymbol{\theta})
-
\lambda_{\mathrm{pen}}
\Delta(\mathbf p,\boldsymbol{\theta}),
\label{eq:ao_penalized_objective}
\end{align}
where $\lambda_{\mathrm{pen}}>0$ denotes the QoS-penalty
coefficient. Accordingly, the merit function used for numerical
refinement is defined as
\begin{align}
\mathcal M(\mathbf p,\boldsymbol{\theta})
=
\begin{cases}
\mathbb U(\mathbf p,\boldsymbol{\theta}),
&
\Delta(\mathbf p,\boldsymbol{\theta})=0,
\\[1mm]
\mathcal F(\mathbf p,\boldsymbol{\theta}),
&
\Delta(\mathbf p,\boldsymbol{\theta})>0.
\end{cases}
\label{eq:ao_merit_function}
\end{align}
Thus, the penalized objective guides the search before QoS feasibility
is attained, whereas the original utility is optimized thereafter.

For fixed RIS phases, the derivative with respect to the power
allocated to active link $i$ is approximated using a central finite
difference:
\begin{align}
\widehat g_{p_i}^{(\ell)}
=
\frac{
\mathcal M
\left(
\mathbf p^{(\ell)}+\delta_p\mathbf e_i,
\boldsymbol{\theta}^{(\ell)}
\right)
-
\mathcal M
\left(
\mathbf p^{(\ell)}-\delta_p\mathbf e_i,
\boldsymbol{\theta}^{(\ell)}
\right)
}{
2\delta_p
},
\label{eq:power_finite_difference}
\end{align}
where $\delta_p>0$ is the perturbation magnitude and $\mathbf e_i$ is
the $i$th canonical vector. A one-sided difference is used whenever a
central perturbation crosses the boundary of the feasible power set.
The projected power candidate is obtained as
\begin{align}
\mathbf p^{\mathrm{cand}}
=
\Pi_{\mathcal P}
\left(
\mathbf p^{(\ell)}
+
\eta_p^{(\ell)}
\widehat{\mathbf g}_{p}^{(\ell)}
\right),
\label{eq:projected_power_candidate}
\end{align}
where $\eta_p^{(\ell)}$ denotes the power-update step size at the
$\ell$th iteration, and $\Pi_{\mathcal P}(\cdot)$ denotes the projection
onto the feasible set
\begin{align}
\mathcal P
=
\left\{
\mathbf p:
p_{k,v}^{b,c}\ge0,\;
\sum_{k\in\mathcal K_v}
\sum_{c\in\mathcal C}
p_{k,v}^{b,c}
\le P_b^{\max},
\ \forall v,b
\right\}.
\label{eq:projected_power_set}
\end{align}
Hence, the nonnegativity and per-BS power-budget constraints are satisfied after every power update.

For fixed transmit powers, the RIS coefficients are parameterized as
\begin{align}
\upsilon_m=e^{\mathrm j\theta_m},
\qquad
\theta_m\in[0,2\pi),
\label{eq:ris_phase_parameterization_ao}
\end{align}
which guarantees $|\upsilon_m|=1$. The derivative with respect to
$\theta_m$ is estimated as
\begin{align}
\widehat g_{\theta_m}^{(\ell)}
=
\frac{
\mathcal M
\left(
\mathbf p^{(\ell+1)},
\boldsymbol{\theta}^{(\ell)}
+\delta_\theta\mathbf e_m
\right)
-
\mathcal M
\left(
\mathbf p^{(\ell+1)},
\boldsymbol{\theta}^{(\ell)}
-\delta_\theta\mathbf e_m
\right)
}{
2\delta_\theta
},
\label{eq:ris_finite_difference}
\end{align}
where $\delta_\theta>0$ is the phase perturbation. The resulting phase
candidate is
\begin{align}
\boldsymbol{\theta}^{\mathrm{cand}}
=
\operatorname{mod}
\left(
\boldsymbol{\theta}^{(\ell)}
+
\eta_\theta^{(\ell)}
\widehat{\mathbf g}_{\theta}^{(\ell)},
2\pi
\right).
\label{eq:projected_ris_candidate}
\end{align}
where $\eta_\theta^{(\ell)}$ denotes the RIS phase-update step size at
the $\ell$th iteration.
Every perturbed and candidate RIS vector is evaluated using the
original system model. In particular, the effective MISO channels and
ZF beamforming vectors are recomputed before evaluating the SINRs,
rates, utility, and QoS shortfall. Consequently, ZF suppression is
adapted to each candidate RIS configuration, while the remaining
inter-cell and cross-VSP interference is explicitly captured.

A QoS-aware rule is used to accept candidate updates. A feasible
candidate is preferred to an infeasible current point. If both points
are feasible, the one with the larger unpenalized utility is retained.
If both are infeasible, they are compared using
$\mathcal F(\mathbf p,\boldsymbol{\theta})$. Hence, once a feasible
iterate is obtained, subsequent accepted updates preserve QoS
feasibility. If an initial candidate is rejected, the corresponding
step size is successively reduced through backtracking.

The power and RIS blocks are alternately updated until neither block
provides an acceptable improvement, the relative merit variation falls
below a prescribed tolerance, or the maximum number of AO iterations
is reached. The resulting method is a projected finite-difference AO
heuristic that preserves the power and unit-modulus constraints and
uses the same exact system evaluation as the DRL environment. It
therefore provides a consistent numerical benchmark, although
convergence to a stationary point, global optimality, and recovery of
a feasible solution from every initialization are not guaranteed.

\subsection{Computational Complexity Analysis}
This section analyzes the computational complexity of the proposed DRL-based spectrum-sharing framework and the EDS benchmark. The computational burden of the DRL approaches mainly comes from deep neural network (DNN) operations, whereas EDS is dominated by the combinatorial enumeration of discrete resource-allocation variables.
In the proposed MISO framework, the action vector includes subchannel scheduling, transmit-power allocation, and RIS phase control. Since ZF beamforming vectors are analytically computed from the instantaneous effective channels, they are not treated as optimization variables. Moreover, the RIS association is fixed by deployment and is therefore excluded from the action space. Hence, the action-space dimension is
\begin{equation}
|\mathcal{A}| =
V B_v K_v C +
V B_v K_v C +
J M_j ,
\end{equation}
where $B_v$ denotes the number of BSs per VSP, $K_v$ is the number of users per VSP, $C$ is the number of subchannels, and $J$ is the number of RISs. The state vector contains the direct and cascaded channel information, user rates, and previous control actions, whose dimension scales approximately as
\begin{equation}
|\mathcal{S}| \propto
V B_v K_v C N_t
+
V B_v J C M_j N_t
+
V K_v J C M_j.
\end{equation}

\subsubsection{Complexity of DRL Training}

Both DDPG and SAC adopt actor--critic architectures, where the main computational cost comes from forward and backward propagation during training. For a mini-batch size $B$, the dominant critic-update complexity scales as
$\mathcal{O}(|\mathcal{S}|\,|\mathcal{A}|\,n)$,
where $n$ denotes the number of neurons per hidden layer. Therefore, for $E$ training episodes and $T$ interaction steps per episode, the overall training complexity scales as
\begin{equation}
\mathcal{O}(E\,T\,B\,|\mathcal{S}|\,|\mathcal{A}|\,n).
\label{eq:complexity_train}
\end{equation}

\subsubsection{Complexity of Online Decision Making}

After training, DRL-based resource allocation only requires a forward pass through the actor network. For a two-hidden-layer neural network, the per-step action-generation complexity is approximately
\begin{equation}
\mathcal{O}(|\mathcal{S}|\,n + n^2 + |\mathcal{A}|\,n),
\end{equation}
which grows polynomially with the state and action dimensions. This avoids the combinatorial search required by exhaustive optimization and enables online decision making in dynamic network scenarios.

\subsubsection{Complexity of EDS With AO-Based Heuristic Refinement}

Let $N_{\mathrm{EDS}}$ denote the number of feasible subchannel-allocation
configurations examined by EDS. Under fixed BS association, its worst-case
scaling is
\begin{align}
N_{\mathrm{EDS}}
=
\mathcal{O}\!\left(
C^{\sum_{v\in\mathcal V}K_v}
\right),
\end{align}
which reduces to $\mathcal{O}(C^{VK})$ when all VSPs serve the same number
$K$ of users. The scheduling constraints generally reduce the number of
configurations evaluated in practice.

Let $N_p$ denote the number of active power variables, $M$ the total number
of RIS elements, and $N_{\mathrm{AO}}$ the number of AO iterations. The
central finite-difference calculations require approximately $2N_p$ and
$2M$ objective evaluations for the power and RIS updates, respectively.
Let $\bar N_{\mathrm{bt}}$ denote the average number of additional
evaluations required by backtracking, and let
$\mathcal C_{\mathrm{eval}}$ denote the complexity of one exact system
evaluation, including the effective-channel, ZF-precoder, SINR, rate, and
utility calculations. The overall complexity is therefore upper-bounded by
\begin{align}
\mathcal{O}\!\left(
\left[
N_{\mathrm{EDS}}
+
N_{\mathrm{AO}}
\left(
2N_p+2M+\bar N_{\mathrm{bt}}
\right)
\right]
\mathcal C_{\mathrm{eval}}
\right).
\label{eq:eds_ao_complexity}
\end{align}
Equivalently, using the worst-case EDS scaling,
\begin{align}
\mathcal{O}\!\left(
\left[
C^{\sum_{v\in\mathcal V}K_v}
+
N_{\mathrm{AO}}
\left(
N_p+M+\bar N_{\mathrm{bt}}
\right)
\right]
\mathcal C_{\mathrm{eval}}
\right),
\end{align}
where constant factors are omitted. Hence, the exponential EDS stage
dominates the overall complexity as the number of users or subchannels
increases, while the subsequent AO refinement grows linearly with the
numbers of active power variables and RIS elements.

\subsubsection{Discussion}
The above analysis shows that DRL shifts most of the computational burden to the offline training stage, while online decision-making only requires polynomial-complexity neural-network inference. In contrast, EDS suffers from exponential complexity due to exhaustive enumeration of discrete scheduling decisions and becomes impractical for large-scale networks. Although DDPG and SAC have the same asymptotic training complexity, SAC incurs a higher practical cost due to its twin critics and entropy regularization. This additional overhead improves exploration, reduces value overestimation, and enhances convergence stability. By comparison, DDPG requires fewer computational resources, making it attractive for embedded and latency-sensitive applications. Therefore, DDPG provides a lower-complexity baseline, whereas SAC achieves superior solution quality for the considered high-dimensional RIS-assisted MISO resource-allocation problem.

\section{Numerical Results}\label{Simulat}

This section evaluates the performance of the proposed DRL-based framework for joint spectrum sharing and RIS configuration. We compare the proposed SAC- and DDPG-based learning approaches under various system configurations. 

\subsection{Channel Model}\label{subsec:channel_model}
 As stated in Section \ref{Sys_Model}
we consider a frequency-selective MISO downlink system. For each subchannel $c\in\mathcal{C}$, the direct BS--UE channel between BS $b$ and user $k$ is modeled as
\begin{align}
\mathbf{h}_{b,k}^{c}
=
\sqrt{\rho_0 d_{b,k}^{-\beta}}\,
\bar{\mathbf{h}}_{b,k}^{c},
\end{align}
where $d_{b,k}$ denotes the Euclidean distance between BS $b$ and user $k$, $\beta$ is the path-loss exponent, and $\rho_0$ is the reference channel gain at a distance of $1$~m. The small-scale fading vector follows independent Rayleigh fading, i.e.,
\begin{align}
\bar{\mathbf{h}}_{b,k}^{c}
\sim
\mathcal{CN}(\mathbf{0},\mathbf{I}_{N_t}),
\quad
\forall\, b,k,c.
\end{align}
For RIS-assisted links, the BS--RIS and RIS--UE channels on subchannel $c$ are modeled as
\begin{align}
\mathbf{G}_{b,j}^{c}
&=
\sqrt{\rho_0 d_{b,j}^{-\beta}}\,
\bar{\mathbf{G}}_{b,j}^{c},\\
\mathbf{r}_{j,k}^{c}
&=
\sqrt{\rho_0 d_{j,k}^{-\beta}}\,
\bar{\mathbf{r}}_{j,k}^{c},
\end{align}
where $d_{b,j}$ and $d_{j,k}$ denote the BS--RIS and RIS--UE distances, respectively. The entries of
$\bar{\mathbf{G}}_{b,j}^{c}\in\mathbb{C}^{M_j\times N_t}$ and
$\bar{\mathbf{r}}_{j,k}^{c}\in\mathbb{C}^{M_j\times1}$
are independently distributed as
$\mathcal{CN}(0,1)$.

The RIS phase shifts are designed with respect to the main carrier frequency and are assumed to be identical across all subchannels. Unless otherwise stated, the channel coefficients are assumed to experience independent Rayleigh fading across different links and subchannels. Throughout the simulations, the path-loss exponent is set to $\beta=2.5$ for all links.
\subsection{Simulation Settings and Benchmarks}\label{sim-setting}
Unless otherwise stated, we consider two VSPs, each operating two BSs equipped with $N_t=4$ antennas and serving $K_v=6$ single-antenna users. Each VSP is allocated $C=4$ subchannels, including $C^{r}=2$ reusable and $C^{d}=2$ dedicated subchannels. At most $L_c=2$ users can be simultaneously scheduled by each BS on one subchannel, which satisfies the ZF feasibility condition $L_c\leq N_t$. The RIS--BS association and RIS coverage are determined by the network topology.

Two RISs are deployed, with one RIS associated with each VSP. Each RIS contains $M_j=10$ reflecting elements and is controlled by its nearest BS. The radius of each VSP service region is $500$~m, and the distance between the two region centers is $800$~m. Users and BSs are randomly deployed within their corresponding regions, while independent channel realizations are generated for different random seeds.

Normalized bandwidth, power, and noise values are adopted, with $B_c=1$, $P_{\max}=1$, and $N_0=10^{-3}$. This normalization preserves the relative SINR and utility comparisons while improving numerical stability. The path-loss exponent is set to $2.5$. The economic parameters are $\lambda^{r}=0.2$, $\lambda^{d}=0.5$, $\psi^{j}=0.3$, and $\alpha_p=0.1$. A minimum-rate requirement of $R_{\mathrm{th}}=0.2$ is imposed, and QoS violations are penalized using $\lambda_{\mathrm{qos}}=50$.

\begin{figure}[t]
	\centering    \includegraphics[width=.95\linewidth]{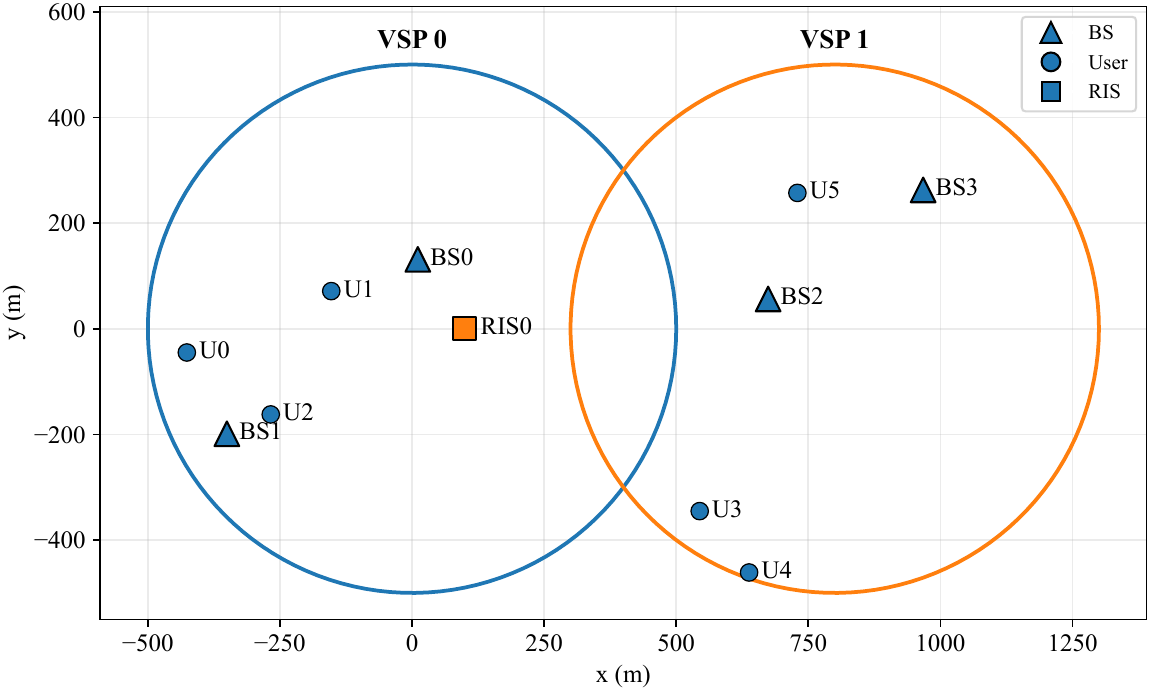}
    \caption{Simulation geometry for a realization with $|\mathcal{B}_v|=2$,
$K_v=3$ and
$J=1$. Users and BSs and RIS are randomly distributed within each  VSP  region.}
	\label{fig-geo}
\end{figure}

The considered benchmarks include DDPG, SAC, and the proposed EDS with AO-based power and RIS phase optimization. Each DRL run contains $10^4$ interaction steps, and the reported convergence curves are averaged over independent random seeds. Both actor and critic networks contain two fully connected hidden layers with 256 neurons and ReLU activations. The actor output is passed through a $\tanh$ function and mapped to the feasible scheduling, power-allocation, and RIS-phase decisions through the projection mechanism described in Section~\ref{sec:DRL}. SAC employs two critic networks and automatic entropy-temperature tuning, whereas DDPG uses a deterministic actor, a single critic, LayerNorm, delayed actor updates, and decaying Gaussian exploration noise. The main simulation and learning parameters are summarized in Tables~\ref{tab:sim_setup} and~\ref{tab:drl_hyper}.
\begin{table}[t]
\centering
\caption{Main Simulation Parameters}
\label{tab:sim_setup}
\footnotesize
\setlength{\tabcolsep}{4pt}
\begin{tabular}{l c}
\hline
\textbf{Parameter} & \textbf{Value} \\
\hline
$(V,|\mathcal B_v|,K_v,N_t)$ & $(2,2,6,4)$ \\
$(C,C^r,C^d,L_c)$ & $(4,2,2,2)$ \\
$(J,M_j)$ & $(2,10)$ \\
$(B_c,N_0,P_{\max})$ & $(1,10^{-3},1)$ \\
Path-loss exponent ($\beta$) & $2.5$ \\
$(R_{\mathrm{th}},\lambda_{\mathrm{qos}})$ & $(0.2,50)$ \\
$(\lambda^r,\lambda^d,\psi^j)$ & $(0.2,0.5,0.3)$ \\
$\alpha_p$ & $0.1$ \\
Region radius / center separation & $500/800$ m \\
\hline
\end{tabular}
\end{table}
\begin{table}[t]
\centering
\caption{Main DRL Hyperparameters}
\label{tab:drl_hyper}
\footnotesize
\setlength{\tabcolsep}{4pt}
\begin{tabular}{l c}
\hline
\textbf{Parameter} & \textbf{Value} \\
\hline
Hidden layers / units & $2/256$ \\
$(\gamma,\tau)$ & $(0.99,5\times10^{-3})$ \\
Batch / replay-buffer size & $256/2\times10^5$ \\
Training steps & $10^4$ \\
SAC learning rates $(\mu_\pi,\mu_Q,\mu_\alpha)$
& $(3,3,3)\times10^{-4}$ \\
DDPG learning rates $(\mu_\pi,\mu_Q)$
& $(1,3)\times10^{-4}$ \\
SAC target entropy & $-|\mathcal A|$ \\
SAC/DDPG gradient updates per step & $2/1$ \\
DDPG policy delay & $2$ \\
Warm-up steps & $1000$ \\
DDPG noise $(\sigma_{\mathrm{init}},\sigma_{\mathrm{final}})$
& $(0.10,0.02)$ \\
\hline
\end{tabular}
\end{table}
Fig.~\ref{fig-geo} illustrates one representative network realization, where users are randomly distributed within their corresponding VSP regions and associated with their nearest BS. The reported results are averaged over independent topology and channel realizations.

\subsection{Performance Evaluation}

\begin{figure}[t]
    \centering
    \includegraphics[width=0.95\linewidth]{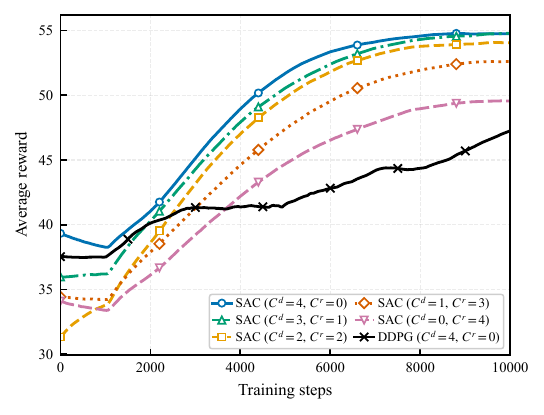}
    \caption{Convergence of SAC for different dedicated and reusable subchannel configurations, and comparison with DDPG for $(C^{d},C^{r})=(4,0)$ ($C=4$ subchannels per VSP).}
    \label{fig4}
\end{figure}

We first investigate the impact of dedicated and reusable subchannels on the proposed RIS-assisted MISO-ZF framework. The total number of subchannels per VSP is fixed to $C=4$, while five spectrum-sharing configurations,
$(C^{d},C^{r})\in
\{(4,0),(3,1),(2,2),(1,3),(0,4)\},$
are considered.
Fig.~\ref{fig4} shows the convergence behavior of SAC under different spectrum-sharing configurations. All cases converge within approximately $8$--$10$k training steps, indicating stable learning. The fully dedicated configuration $(C^{d},C^{r})=(4,0)$ achieves the highest reward, converging to approximately $55$, since inter-VSP interference is eliminated and the joint optimization of scheduling, power allocation, and RIS phase shifts can fully exploit the available spatial degrees of freedom.

As the number of reusable subchannels increases, the reward gradually decreases due to stronger inter-VSP interference. Nevertheless, the degradation is relatively small. Even the fully reusable case $(C^{d},C^{r})=(0,4)$ converges to approximately $50$, corresponding to only about a $10\%$ reduction compared with the fully dedicated configuration. This demonstrates that the proposed RIS-assisted MISO-ZF framework effectively suppresses interference through ZF beamforming and adaptive RIS optimization, enabling efficient spectrum reuse with only a limited utility loss.

Fig.~\ref{fig4} also compares SAC and DDPG for the fully dedicated configuration $(C^{d},C^{r})=(4,0)$. SAC consistently converges faster and achieves a higher final reward (approximately $55$) than DDPG (approximately $47$). This improvement is mainly attributed to SAC's entropy-regularized exploration and twin-critic architecture, which provide more stable policy optimization in the high-dimensional joint resource allocation problem.
Overall, the results indicate that dedicated subchannels maximize network utility, whereas moderate spectrum reuse, particularly $(C^{d},C^{r})=(3,1)$ and $(2,2)$, achieves comparable performance while improving spectrum utilization.

\begin{figure}[t]
    \centering
    \includegraphics[width=0.95\linewidth]{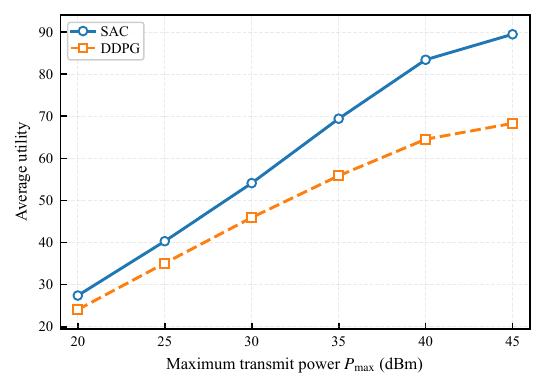}
    \caption{Average utility versus the maximum transmit power for SAC and DDPG with $(C^{d},C^{r})=(2,2)$.}
    \label{fig5}
\end{figure}

Fig.~\ref{fig5} compares the utility achieved by SAC and DDPG as the maximum transmit power varies from 20 to 45 dBm. The remaining simulation parameters follow Tables~\ref{tab:sim_setup} and~\ref{tab:drl_hyper}. SAC consistently outperforms DDPG over the entire power range. At low transmit-power levels, particularly below 30 dBm, satisfying the QoS constraints is difficult for both methods, resulting in a relatively small performance gap. As the available power increases, the QoS constraints become easier to satisfy, and SAC achieves noticeably higher utility. This improvement is attributed to the entropy-regularized objective of SAC, which promotes broader exploration and increases the likelihood of identifying better resource-allocation policies.

\begin{figure}[t]
    \centering
    \includegraphics[width=0.95\linewidth]{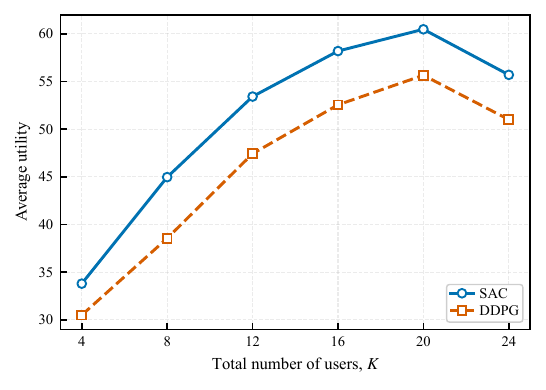}
    \caption{Average utility versus total number of users $K$.}
    \label{fig6}
\end{figure}

Fig.~\ref{fig6} illustrates the average utility as a function of the
total number of users, $K$. The utility initially increases with $K$
because a larger number of users provides more opportunities to improve
the aggregate sum rate. However, when $K$ exceeds 20, the average
utility decreases. In this denser regime, the available spectrum and
transmit-power resources must be shared among more users, while the
intra-VSP and inter-VSP interference levels also increase. As a result,
the marginal sum-rate gain is outweighed by the higher resource
contention and interference. Across all considered user densities, SAC
consistently outperforms DDPG, confirming its greater robustness in
larger and more strongly coupled action spaces.

\begin{figure}[t]
    \centering
    \includegraphics[width=0.95\linewidth]{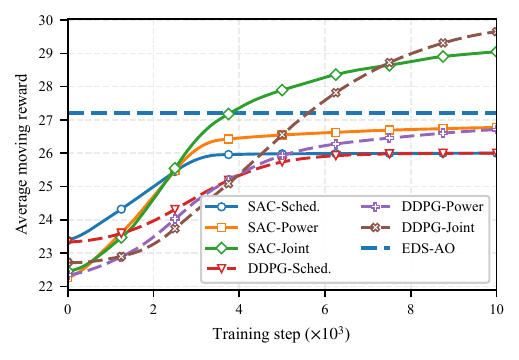}
   \caption{Convergence comparison of the SAC- and DDPG-based ablation variants with the EDS-AO heuristic benchmark.}
    \label{fig7}
\end{figure}

Fig.~\ref{fig7} compares the convergence behaviors of the proposed
SAC- and DDPG-based resource-allocation strategies together with the
EDS-AO heuristic benchmark. To keep the computational complexity of
the exhaustive search manageable, this study employs a reduced-scale
system with $V=2$ VSPs, $B_v=1$ BS per VSP, $K_v=2$ users per VSP,
$C=2$ subchannels, $J=1$ RIS comprising $M_j=4$ reflecting elements,
$N_t=2$ BS antennas, $L_c=1$ dedicated subchannel, and
$C_r=1$ reusable subchannel. Consequently, the action space is
significantly smaller than that used in the main simulation setup.
For both DRL algorithms, three optimization configurations are
considered. The \emph{Scheduling} variant optimizes only the
subchannel allocation while employing equal power allocation and fixed
RIS phases. The \emph{Power} variant jointly optimizes the
subchannel allocation and transmit-power allocation with fixed RIS
phases. Finally, the \emph{Joint} variant simultaneously optimizes
the subchannel allocation, transmit-power allocation, and RIS phase
shifts.

As expected, the Scheduling variants converge to the lowest reward,
indicating that optimizing only the discrete scheduling decisions
provides limited performance gains. Incorporating transmit-power
allocation further improves the reward by adapting the continuous
resource allocation to the selected scheduling configuration.
Simultaneously optimizing scheduling, power allocation, and RIS phase
shifts achieves the largest performance improvement, demonstrating the
strong coupling among these optimization variables.

The EDS-AO heuristic provides a stronger benchmark than the
scheduling-only and power-only solutions by refining the
continuous variables after exhaustive scheduling. Nevertheless, its
final reward remains below those achieved by the converged DRL agents,
since the discrete scheduling decisions are fixed before the AO
refinement, whereas the DRL policies jointly learn the coupled
scheduling, power-allocation, and RIS optimization policy in an
end-to-end manner.
Although DDPG-Joint achieves a slightly higher final reward under the reduced-scale configuration, SAC consistently outperforms DDPG in the larger-scale experiments, demonstrating better scalability as the optimization problem becomes more complex.

To further evaluate the proposed framework under a larger and more challenging scenario, we consider a system with $V=3$ VSPs, each serving $K_v=10$ users over $C=8$ subchannels, $L_c=4$ and $C_r=4$ reusable subchannels. Three RISs are deployed, each equipped with $M_j=16$ reflecting elements, resulting in a total of $48$ RIS elements, while each BS employs $N_t=4$ transmit antennas. This configuration yields state and action dimensions of $38\,664$ and $528$, respectively, and therefore provides a challenging high-dimensional setting for evaluating the proposed DRL framework.
Table~\ref{tab:large_scale_scalability} compares SAC and DDPG under this configuration. The experiments were conducted on a workstation equipped with a 13th-generation Intel Core i9-13900K CPU operating at 3.00~GHz and 64~GB of RAM, using CPU-only execution without GPU acceleration. 
Besides the final moving-average reward, which reflects the convergence performance after training, the table reports the total training time for $20\,000$ environment interactions, the average actor inference latency for a single decision, the peak RAM consumption during training, and the replay-buffer memory allocation. Although SAC requires approximately 13\% more training time than DDPG, it achieves approximately 45\% higher final moving-average reward due to its entropy-regularized twin-critic architecture. Moreover, the inference latency remains below $1$~ms for both algorithms, making online deployment feasible. The peak RAM usage of both methods is comparable (approximately $6.8$--$7.0$~GB), while the replay-buffer memory is identical because both algorithms employ the same replay-buffer size and environment representation. These results demonstrate that the performance improvement achieved by SAC is obtained with only a modest increase in computational cost.
\begin{table}[t]
\centering
\caption{Performance and computational comparison}
\label{tab:large_scale_scalability}
\resizebox{\columnwidth}{!}{%
\begin{tabular}{lccc}
\toprule
\textbf{Metric} & \textbf{Unit} & \textbf{SAC} & \textbf{DDPG} \\
\midrule
Final moving-average reward & -- & $45.573$  & $31.001 $ \\
Training time & min & $58.58$ & $51.85 $ \\
Inference time & ms & $0.9476$ & $0.7208$ \\
Peak RAM usage & MB & $6969.0 $ & $6782.7$  \\
Replay-buffer memory & MB & $5940.1 $ & $5940.1 $ \\
\bottomrule
\end{tabular}%
}
\end{table}

\subsection{Impact of DRL Hyperparameters}
\label{subsec:hyperparam}

This subsection evaluates the sensitivity of the proposed DRL algorithms to the learning rate $\mu$ and mini-batch size $B$. The actor and critic learning rates are set equal, i.e., $\mu=\mu_{\pi}=\mu_{Q}$. The experiments follow the simulation setup in Section~\ref{sim-setting}. Each curve is obtained using the same random seed and smoothed with a moving average window of 500 training steps.

\begin{figure}[t]
    \centering
    \subfloat[DDPG]{\includegraphics[width=0.49\linewidth]   {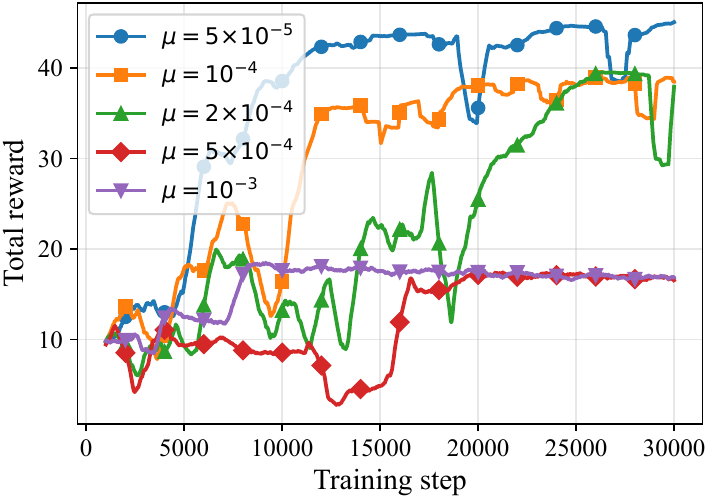}}    
    \hfill
    \subfloat[SAC]{\includegraphics[width=0.49\linewidth]{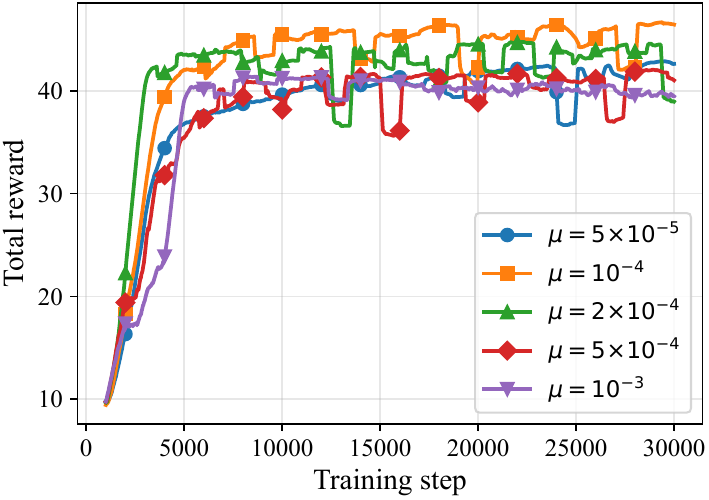}}
    \caption{Impact of the learning rate $\mu$ on the training performance under the spectrum-sharing system described in Section~\ref{sim-setting}.}
    \label{fig:lr_sweep}
\end{figure}

Fig.~\ref{fig:lr_sweep} compares the learning-rate sensitivity of SAC and DDPG. SAC exhibits limited sensitivity to the learning rate, with final objective values remaining within approximately 40--48 across all tested settings. In contrast, DDPG is substantially more sensitive to the learning-rate selection. Although its best configuration achieves a performance comparable to SAC, increasing the learning rate to $\mu\ge5\times10^{-4}$ reduces the final objective by more than 20 units, indicating unstable training under aggressive gradient updates.

\begin{figure}[t]
    \centering
    \subfloat[DDPG]{\includegraphics[width=0.49\linewidth]{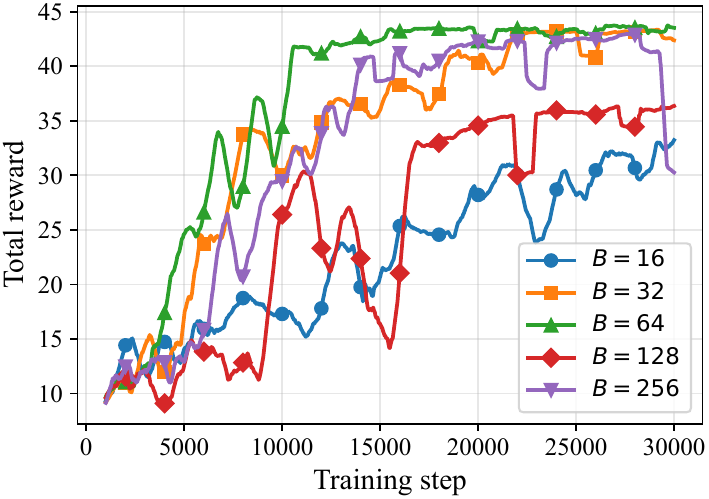}}   
    \hfill
    \subfloat[SAC]{\includegraphics[width=0.49\linewidth]{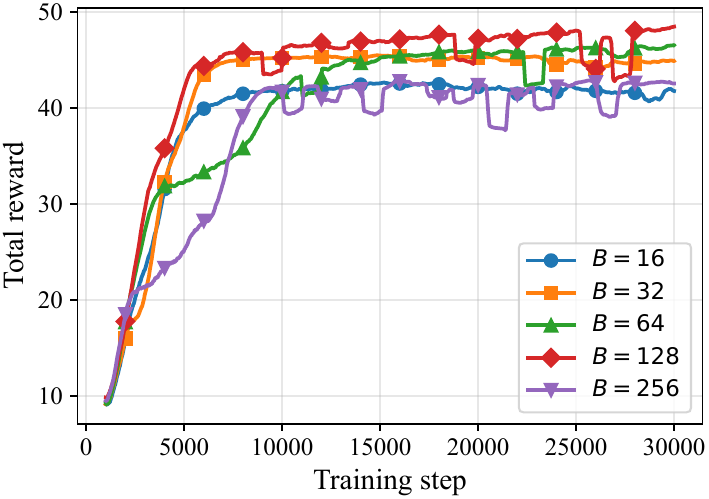}}
    \caption{Impact of the mini-batch size $B$ on the training performance under the spectrum-sharing system described in Section~\ref{sim-setting}.}
    \label{fig:batch_sweep}
\end{figure}

Fig.~\ref{fig:batch_sweep} illustrates the influence of the mini-batch size. SAC achieves objective values within a narrow range for all tested batch sizes, whereas DDPG experiences a noticeable performance reduction when small mini-batches are employed. For example, the objective obtained with $B=16$ is approximately 10 units lower than that achieved using the best-performing batch size. With their respective best batch-size settings, SAC again outperforms DDPG by approximately $9$--$10\%$.

Overall, the hyperparameter study indicates that SAC provides more consistent performance over a broad range of learning rates and batch sizes. The observed improvement is primarily attributed to the entropy-regularized objective and twin-critic architecture, which improve value estimation and stabilize policy optimization.

\section{Conclusion}\label{conc}
This paper investigated dynamic spectrum sharing for RIS-assisted LHQWNs operating within an MNO–VSP ecosystem. The joint optimization of subchannel allocation, transmit power control, and RIS phase configuration was formulated as a utility maximization problem under spectrum leasing costs, RIS deployment costs, and QoS constraints. Due to the resulting mixed-integer nonlinear structure, the problem was modeled as an MDP and solved using DRL techniques.
Two actor--critic algorithms, DDPG and SAC, were developed and evaluated. Numerical results demonstrated that the proposed SAC-based solution consistently outperforms DDPG in terms of convergence speed, training stability, and achievable utility, particularly in larger-scale and more strongly coupled scenarios. The proposed framework also exhibits low inference latency and practical memory requirements under the considered large-scale system configuration, demonstrating its computational efficiency.
The results further confirmed the performance benefits of RIS deployment. Since the RIS leasing cost is fixed, optimizing RIS phase configurations significantly enhances effective channel gains and overall VSP utility. Furthermore, the proposed joint optimization framework consistently benefits from increasing transmit power and effectively adapts to different network sizes and spectrum-sharing configurations. The ablation study further showed that SAC provides more robust performance than DDPG as the optimization problem becomes increasingly coupled.

Overall, the proposed framework provides an effective and scalable solution for mixed discrete–continuous resource optimization in RIS-assisted spectrum sharing environments.
In this work, QoS was captured through a minimum-rate constraint. As future work, the proposed framework can be extended to incorporate latency- and reliability-aware QoS requirements, multi-RIS cooperative deployments, and dynamic environments with time-varying CSI and user mobility.

\bibliographystyle{ieeetr} 
\bibliography{references1}

\end{document}